\documentstyle[epsfig]{proceed}

\def\891{NGC~891}
\def\4013{NGC~4013}

\def\farcs{\hbox{$.\!\!^{\prime\prime}$}}
\def\z{$z$}
\def\wiyn{WIYN}
\def\halpha{H$\alpha$}
\def\cmsq{cm$^{-2}$}
\def\msun{M$_\odot$}
\def\HII{H$\;${\small\rm II}\relax}
\def\Nh{$N_{\rm H}$}

\def\lesssim{\mathrel{\hbox{\rlap{\hbox{\lower4pt\hbox{$\sim$}}}\hbox{$<$}}}}
\def\gtrsim{\mathrel{\hbox{\rlap{\hbox{\lower4pt\hbox{$\sim$}}}\hbox{$>$}}}}
\let\la=\lesssim
\let\ga=\gtrsim

\def\etal{{\em et al.}}

\title[Extraplanar Dust and Star Formation]
{Extraplanar Dust and Star Formation in Nearby Edge-On Galaxies}

\author[J.C. Howk \& B.D. Savage]
{J. Christopher Howk\email{howk@astro.wisc.edu} \and{} 
Blair D. Savage\email{savage@astro.wisc.edu}}

\institute{University of Wisconsin-Madison}

\begin{document}

\maketitle

\abstract{We present high-resolution ($\lesssim0\farcs7$) ground-based
images of the edge-on spiral galaxies NGC~891 ($D\sim9.5$ Mpc) and
NGC~4013 ($D\sim17$ Mpc) acquired with the \wiyn\ 3.5-m telescope.
These BVI+\halpha\ images reveal complex webs of dusty interstellar
material far above the midplanes of both galaxies ($0.5 \lesssim |z|
\lesssim 2$ kpc) as discussed previously by Howk \& Savage (1997).
The dusty high-\z\ clouds, visible in absorption against the
background stellar light of the galaxies, have widths 50--100 pc and
lengths 100--400 pc. An analysis of their absorbing properties
suggests they have $A_V \gtrsim 0.5 - 2.0$.  If Milky Way gas-to-dust
relationships are appropriate, then these structures have gaseous
column densities $N_{\rm H} \gtrsim 10^{21}$ \cmsq, with very large
masses ($\gtrsim 10^5 - 10^6$ \msun) and gravitational potential
energies ($\gtrsim 10^{51} - 10^{52}$ ergs relative to $z=0$).  The
estimated column densities suggest molecular gas may be present, and
with the estimated masses allows for the possibility of star formation
in these dusty clouds.  Recent star formation is the likely cause of
the discrete \HII\ regions, in some cases associated with relatively
blue continuum sources, observed at heights $0.6 \lesssim |z| \lesssim
0.8$ kpc above the disks of these galaxies.  The presence of
early-type stars at high-\z\ in these galaxies may be related to the
extraplanar dust structures seen in our images.}

\section{Introduction}

Recent studies have shown many of the phases of the interstellar
medium (ISM) found in the thin interstellar disks of spiral galaxies
are also present in their extended gaseous thick disks (e.g., Dettmar,
these proceedings).  Galaxies which are vigorously forming stars may
also be expelling a substantial amount of hot enriched gas as young
massive stars evolve and die as supernovae, and the existence of thick
disk gas seems to be correlated with the star formation rate of the
underlying disk.  The effects of supernova explosions are not,
however, limited to the hot gas, but rather massive stars may be able
to expel parcels of ambient gas to great heights above the planes of
spiral galaxies.

Whatever the precise mechanism for supplying gas to the thick disk, it
is clear that warm and hot material are found in this interface region
between the thin disks and the more extended gaseous halos of
galaxies.  What, however, are the conditions of this material?  How is
the thermal and multi-phase balance of this material different than
that in the disk?  Are dust grains and molecules present at large
distance from the plane?

We have undertaken an imaging study of several nearby edge-on spiral
galaxies with the WIYN 3.5-m telescope to search for and characterize
gas and dust structures in the disk-halo interface of these systems.
In this contribution we present images of two of the galaxies in our
study: \891\ and \4013, extending our earlier imaging work on \891\
(Howk \& Savage 1997; hereafter \cite{hs97}).  Here we briefly
describe the conclusions we have drawn from our images, particularly
regarding the physical properties of dusty clouds in the interstellar
thick disks of \891\ and \4013.  We also comment on evidence in our
images for the presence of early-type stars far above the planes of
spiral galaxies.

\section{Dusty Thick Disks in Edge-On Spirals}

Using the \wiyn\ 3.5-m telescope on Kitt Peak we have obtained
high-resolution ($\la 0\farcs7$) BVI+\halpha\ images of several nearby
edge-on galaxies.  Fig. \ref{fig:unsharp} presents unsharp-masked
versions of the V-band and B-band images for \891\ and \4013,
respectively.  These images reveal extensive amounts of
highly-structured dust seen in absorption against the background
stellar light of these galaxies to heights $z \la 2$ kpc from their
miplanes.

An analysis of the absorbing properties of the dust in the BVI
bandpasses allows us to estimate the total visual extinction, $A_V$,
through the observed dust features (e.g., \cite{hs97}).  In
Fig. \ref{fig:unsharp} we have marked several individual dust
structures in each galaxy; Table \ref{tab:dust} summarizes the
properties of these structures.  Our analysis neglects the effects of
scattering by grains, but this omission is such that our derived
values of $A_V$ are lower limits.  The entries given for \Nh, mass,
and gravitational potential energy in Table \ref{tab:dust} are derived
assuming Milky Way gas-to-dust relationships (e.g., \cite{bsd78}).

Our analysis suggests that the observed dust features in these two
galaxies are quite massive ($\ga~10^5 - 10^6$ \msun) with very large
energy requirements if ejected from $z=0$ ($\ga~10^{52} - 10^{53}$
ergs).  Though these dust clouds may be slightly larger (typically
50--100 pc$\times$100--400 pc), their masses are quite similar to the
low-\z\ molecular cloud complexes in the inner Milky Way
(\cite{dame86}).  We note that for gas column densities like those
estimated for these structures ($N_{\rm H} \ga 10^{21}$), the fraction
of hydrogen found in molecular form in diffuse Milky Way disk clouds
is $f \sim 25\%$ (\cite{savage77}).  Though we do not fully understand
the physical properties of these high-\z\ clouds, if the conditions
are not too different than for diffuse material in the disk of the
Milky Way, particularly the radiation field, these dusty clouds may
contain a significant amount of molecular material.  Searches for CO
emission from these clouds would be of great interest.

\begin{table}
\caption{Dust cloud properties in NGC~891 and NGC~4013 \label{tab:dust}}
\begin{center}
\begin{tabular}{crcccccrcccc}
\hline \hline
 & \multicolumn{5}{c}{NGC 891} & & \multicolumn{5}{c}{NGC 4013} \\
\cline{2-6} \cline{8-12}
Cloud$^1$ & 
 \multicolumn{1}{c}{\z} & $A_V$ & $N_{\rm H}\,^2$ & Mass$^2$ 
	& Energy$^{2,3}$ & &
 \multicolumn{1}{c}{\z} & $A_V$ & $N_{\rm H}\,^2$ & Mass$^2$ 
	& Energy$^{2,3}$ \\
 & \multicolumn{1}{c}{[pc]} & [mag.] & [\cmsq] & [$10^5$ \msun] 
	& [$10^{51}$ ergs] & &
   \multicolumn{1}{c}{[pc]} & [mag.] & [\cmsq] & [$10^5$ \msun] 
	& [$10^{51}$ ergs] \\
\hline
1 & 600 & 1.8 & $3\times10^{21}$ &
        40 & 130 & & 
   1000 & 0.5 & $1\times10^{21}$ &
        9 & 70 \\
2 & 1450 & 0.8 & $1\times10^{21}$ &
        2 & 25 & & 
     750 & 0.8 & $2\times10^{21}$ &
        2 & 7 \\
3 & 600 & 0.7 & $1\times10^{21}$ &
        8 & 40 & & 
    640 & 1.3 & $3\times10^{21}$ &
        9 & 30 \\
4 & 950 & 0.7 & $1\times10^{21}$ &
        3 & 25 & & 
    720 & 0.9 & $2\times10^{21}$ &
        5 & 20 \\
5 & 1350 & 0.7 & $1\times10^{21}$ &
        2 & 30 & & 
     530 & 0.5 & $1\times10^{21}$ &
        2 & 5 \\
\hline 
& \multicolumn{11}{l}{\footnotesize $^1$ -- Identification number in
 Fig. \ref{fig:unsharp} for individual dust clouds in both
 galaxies.}\\
 & \multicolumn{11}{l}{\footnotesize $^2$ -- These quantities assume
	Galactic gas-to-dust relationships.}\\
 & \multicolumn{11}{l}{\footnotesize $^3$ -- Gravitational potential
energy relative to the midplane.}  
\end{tabular}
\end{center}
\end{table}

What are the origins of the dusty clouds found at high-\z\ in edge-on
spirals such as \891\ and \4013?  The observed numbers and properties
of dust structures seem to be roughly symmetric about the planes of
these galaxies, suggesting that they are not associated with warped
gas layers.  \cite{hs97} discuss several mechanisms for expelling
dusty material from the thin disks of spirals.  Among the more
promising of these scenarios are expulsion via (magneto)hydrodynamical
flows (\cite{ni89}) and/or radiation pressure (\cite{f93}).  We
find that some of the observed dust structures are associated with
sites of vigorous star formation in the underlying disk.  Indeed some
of the morphologies (e.g., shell or cone-like structures) suggest
these dust features may be tracing the expulsion of matter in
supernova-driven outflows.  Generally we find that such dust
structures have closely associated \halpha\ emission.  Given the
morphologies and associated \halpha\ emission, it is very likely that
massive stars and supernovae are playing a role in shaping some of the
observed structures.  However, most dust structures seem not to have
bright associated \halpha\ and their morphologies are less easily
associated with the immediate effects of star formation in the disk.
Other scenarios must also be considered.  For example, it may be
possible to form high-\z\ clouds through thermal instabilities in hot
extraplanar gas.  One of the most significant implications of our
images is that whatever processes provide for interstellar thick disk
material, the material at high-\z\ in these galaxies must include a
significant amount of dust.  Thus, if these structures are formed as
material is lifted away from the thin disk, the mechanisms responsible
for the expulsion cannot be violent enough to destroy the associated
dust.

\section{High-\z\ Star Formation}

In the course of our studies of the dusty ISM at large \z -heights, we
have identified regions that contain early-type stars far from the
thin disks of several edge-on spirals.  Fig. \ref{fig:highzstars}
shows B-band gray-scale images with \halpha\ contours overlayed for
two such regions.  In both \891\ and \4013\ we find unresolved knots
of \halpha\ emission at \z -heights of 0.6--1 kpc.  We tentatively
identify these knots of emission as \HII\ regions, though this
identification should be confirmed spectroscopically.  

In \891, the closer of the two galaxies studied here, we also detect
continuum emission in the BVRI bandpasses from several sources
associated with \halpha\ emission at $|z| \ga 0.5$ kpc.  Though our
photometry is not yet complete, the high-\z\ continuum sources in
\891\ seem to require the equivalent of more than 5 main sequence
O-stars to produce the observed light.  While single early-type stars
may be ejected from the thin disk during their lifetimes, most
mechanisms for such ejections (e.g., supernova kicks, dynamical
ejection from clusters, etc.) seem incapable of explaining groups of
stars at such large distances from the midplane.  If these regions do
indeed contain multiple early-type stars, it suggests that a mode of
star formation is operable beyond the thin (inter)stellar disks of
spiral galaxies, possibly associated with the massive dust clouds seen
in our images.

\vspace{0.05in} 
We thank the many people who have contributed to the
planning and construction of the WIYN Observatory and in its
scientific optimization and acknowledge major support from the higher
administration of the University of Wisconsin-Madison for our
department's participation in this project.

\begin{figure}[ht]
\caption{Unsharp-masked views of NGC~891 (top; V-band) and NGC~4013
(bottom; B-band).  These images were taken with the \wiyn\ 3.5-m
telescope and have seeing-limited resolutions of $0\farcs7$ or
$\sim32$ pc for NGC~891 and $0\farcs6$ or $\sim50$ pc for NGC~4013).
These images show hundreds of absorbing dust structures with heights
$0.5 \la z \la 2$ kpc from the midplane.  Several dust structures are
identified in these images, and the physical properties of these
structures are summarized in Table \ref{tab:dust}.  A 1 kpc distance
scale is shown on the left side of each image.}
\label{fig:unsharp}
\end{figure}

\begin{figure}
\centering
\caption{Gray-scale B-band data for \891\ (left) and \4013\ (right)
with \halpha\ contours overlayed.  Prominent \HII\ regions far above
the plane are identified in both galaxies.  These \HII\ regions and,
in the case of \891, associated continuum sources imply the presence
of groups of early-type stars at high-\z, strongly suggesting that
stars may be formed from interstellar thick disk material.}
\label{fig:highzstars}
\end{figure}

\end{document}